\documentclass[lettersize,journal]{IEEEtran}
\usepackage{amsmath,amsfonts}
\usepackage{algorithmic}
\usepackage{algorithm}
\usepackage{array}
\usepackage[caption=false,font=normalsize,labelfont=sf,textfont=sf]{subfig}
\usepackage{textcomp}
\usepackage{stfloats}
\usepackage{url}
\usepackage{verbatim}
\usepackage{graphicx}
\usepackage{cite}
\usepackage{xcolor}
\usepackage[framemethod=TikZ]{mdframed}

\usepackage{subfig}

\usepackage{tabularx}
\usepackage{threeparttable}
\usepackage{algorithm}
\usepackage{algorithmic}
\usepackage{booktabs}

\usepackage{array}
\usepackage{multirow}

\newcolumntype{P}[1]{>{\centering\arraybackslash}p{#1}}
\newcolumntype{M}[1]{>{\centering\arraybackslash}m{#1}}

\begin{document}

\title{An Open API Architecture to Discover the Trustworthy Explanation of Cloud AI Services}


\author{
	\IEEEauthorblockN{
		Zerui Wang,
        Yan Liu,
		Jun Huang 
  }
 
    \IEEEauthorblockA{\textit{Department of Electrical and Computer Engineering} \\
    \textit{Concordia University}\\
    Montréal, Québec, Canada \\
    \{zerui.wang, yan.liu, jun.huang\}@concordia.ca}
  }

\markboth{Journal of \LaTeX\ Class Files,~Vol.~14, No.~8, August~2021}%
{Shell \MakeLowercase{\textit{et al.}}: A Sample Article Using IEEEtran.cls for IEEE Journals}

\maketitle

\thispagestyle{plain}
\pagestyle{plain}

\begin{abstract}
This paper presents the design of an open-API-based explainable AI (XAI) service to provide feature contribution explanations for cloud AI services.
Cloud AI services are widely used to develop domain-specific applications with precise learning metrics.
However, the underlying cloud AI services remain opaque on how the model produces the prediction. 
We argue that XAI operations are accessible as open APIs to enable the consolidation of the XAI operations into the cloud AI services assessment. 
We propose a design using a microservice architecture that offers feature contribution explanations for cloud AI services without unfolding the network structure of the cloud models. 
We can also utilize this architecture to evaluate the model performance and XAI consistency metrics showing cloud AI services' trustworthiness. 
We collect provenance data from operational pipelines to enable reproducibility within the XAI service. 
Furthermore, we present the discovery scenarios for the experimental tests regarding model performance and XAI consistency metrics for the leading cloud vision AI services. 
The results confirm that the architecture, based on open APIs, is cloud-agnostic.
Additionally, data augmentations result in measurable improvements in XAI consistency metrics for cloud AI services.
\end{abstract}


\section{Introduction}
\label{sec:introduction}

\IEEEPARstart{A}{rtificial} Intelligence (AI) as a service is a rapidly expanding technology paradigm.
The AI market is expected to grow as more companies adopt AI technology to remain competitive.
Major cloud providers, including Amazon Web Services~\cite{aws-ai-services}, Microsoft Azure~\cite{azure-cognitive-services}, Google Cloud Platform~\cite{google_cloud_vertex_ai}, Alibaba Cloud~\cite{AlibabaCloud}, Oracle Cloud, IBM Cloud, and Salesforce Service Cloud, offer AI as a service. This enables customers to develop and deploy AI models using cloud-based platforms.

Cloud AI services commonly achieve learning accuracy by using standard metrics such as precision, recall, and F1 score~\cite{aws-ai-services,azure-cognitive-services,google_cloud_vertex_ai}.
However, certain services such as Alibaba Cloud~\cite{AlibabaCloud} may not explicitly provide these performance metrics. 
Additionally, a recent study \cite{cummaudo2019losing} conducts a detailed investigation into the maintenance of AI services, focusing on computer vision. 
This research uncovers inconsistencies and evolution risks in AI services. 

The explainable AI (XAI) aims to develop models and methods that enable human users to comprehend, trust, and manage AI models~\cite{DARPA}. 
A study~\cite{meske2020transparency} discusses the role of XAI in computer vision-based decisions. 
They emphasize that XAI can promote trust in AI computer vision systems through improved understanding and prediction. 
Another work~\cite{paper1} presents the XAI criterion that refines the functionality objectives for XAI methods. 
One criterion involves the analysis of feature influence and feature causality. 

XAI is increasingly adopted in applications that require transparency, fairness, and trustworthiness in decision-making, particularly in sensitive domains~\cite{amann2020explainability}.
The limitation of the XAI practices is the existing XAI techniques developed are often tailored to specific types of models or cases~\cite{yang2022omnixai}, which makes the XAI practices less reusable and versatile to other applications. 
Meanwhile, numerous cloud AI services have been provided by cloud platforms to support general applications across domains~\cite{aws-ai-services,azure-cognitive-services,google_cloud_vertex_ai,AlibabaCloud}. 
Together, the trend underscores the need to integrate XAI methods with cloud AI services to foster trust in cloud-based AI applications.


XAI demands that activities conducted during an XAI process be traceable and reproducible~\cite{zafar2017trustworthy}. 
Ensuring that the data utilized in the AI models and XAI methods are trustworthy becomes important. 
Addressing data provenance in XAI operations is essential for guaranteeing that the generated explanations are reliable, verifiable, and consistently reproducible across diverse settings. 

Despite the numerous XAI frameworks available, a noticeable gap exists among essential components \cite{yang2022omnixai} including data processing, methods configuration, and evaluation metrics. 
These components collectively form complex pipelines. 
This observation motivates our proposal: a design by a cloud-native paradigm based on the microservice architecture. 
This architectural style benefits from its capacity for the independent deployment of diverse components, streamlined communication via RESTful APIs, and a built-in adaptability that accommodates the introduction of new XAI methods, substitutions of AI models, and adjustments in pipeline configurations.
Furthermore, our proposed architecture offers precise record capabilities. This ensures that the provenance of every XAI operation is transparent, facilitating the reproduction of XAI tasks or entire workflows.

For black-box models or cloud-based AI services, the task of revealing the internal structures of AI models becomes unfeasible, especially for those AI services that encapsulate models behind standard RESTful APIs. 
We address this challenge by drawing inspiration from XAI methods that focus on feature influence and causality, for example, SHAP~\cite{lundberg2017unified}. 
Besides, we propose a method that approximates the black-box AI model with a custom-built model and computes the feature contribution values, providing interpretable insights even from opaque AI models.
Confronted by these challenges, our work is directed towards the following research questions.\\
\begin{itemize}
    \item \textbf{\textit{RQ1: How to obtain and evaluate XAI results without unfolding the cloud AI service model structure?}}

    We investigate the cloud AI services and XAI methods in Section \ref{sec:Cloud-based} and \ref{sec:reviewXAImethods}.
    This enables us to understand the communication between cloud AI services and the specific requirements for XAI methods. Subsequently, we briefly summarize the applicable XAI methods in the taxonomy, Section \ref{sec:taxonomy}. We also seek the packaged XAI frameworks listed in Section \ref{sec:framework}. However, most frameworks are not explicitly compatible with cloud AI services.
    Therefore, we propose workflow as Figure \ref{flow-chart} that integrates Cloud AI with Post hoc XAI, expressed in Section \ref{integratingcloud}. Ultimately, scenarios one and two in Section \ref{sec:scenario1} and \ref{sec:scenario2} compute and evaluate the XAI results from integrating three major cloud AI services. 

    \item \textbf{\textit{RQ2: What are the essential components required for XAI service architecture to deliver feature contribution explanations for models?}}

    To implement XAI within a service-oriented framework, Section \ref{sec:Design} delineates the key architectural components critical for integration with existing cloud-based AI services. Following this, Section \ref{sec:Discovery} presents four illustrative scenarios using the designed XAI service architecture to explore typical discovery situations.

    \item \textbf{\textit{RQ3: How to collect XAI provenance data from operations to ensure traceability within the XAI service?}}

    Referring to the related works in Section \ref{sec:relatedprovenance}, we notice that the provenance data is necessary for XAI operations.
    Referring to the key components in XAI operations, we provide a graph format design for the XAI provenance data. 
    Section \ref{sec:provenance} introduces how to automatically collect the provenance data from various XAI operations within the XAI service. By retrieving the provenance data, we can identify differences and edit configurations to the XAI operations. In section \ref{sec:scenario3}, scenario three, we showcase a scenario that optimizes the model by modifying and executing reproduction. This scenario leads to improvements in both model performance and the XAI evaluation metrics. With the operations traceable and reproducible, we present the cloud-agnostic reproduction in scenario four, section \ref{sec:scenario4}.
     
\end{itemize}

In this work, we propose an innovative XAI service architecture specifically designed to feature contribution explanations, illustrated through a showcase scenario drawn from computer vision cloud AI services. 
This method involves the utilization of approximation models to generate images, emphasizing the most contributing features. These masked images then act as inputs to create the AI services' predictions. We calculate the prediction changes value between the original and masked images. 
Leveraging these prediction changes, we compute a comprehensive explanation summary for the AI services, providing a transparent overview. 





The main contributions are summarized as follows:
\begin{itemize}

    \item \textbf{Design cloud-platform-independent XAI service framework.} The open API architecture is independent of the cloud-specific AI service. The architecture accommodates first-class entities in the XAI process as unified micro-services. The communication is open API-based, thus encapsulating the variance of models, XAI methods, and inputs and outputs from feature engineering.  
    \item \textbf{Provide explanations across multiple cloud AI services.} Based on the definition of the XAI consistency metric, we derive an explanation summary cross-validated on multiple clouds to observe both the learning performance of AI services and data augmentation effects. 
    \item \textbf{Reproduce XAI operations through configure-and-rerun.} The configuration of services is the receipt of composing an end-to-end explanation workflow. By reserving the configuration of each service given a workflow definition, we accumulate the provenance of how each explanation is produced. Through the coordination center of the XAI framework, we can rerun the XAI workflow to reproduce the explanation. 

\end{itemize}

We demonstrate the XAI service architecture with four discovery scenarios in Section \ref{sec:Discovery}, including (1) Cloud AI performance evaluation, (2) XAI consistency evaluation, (3) Probing of data augmentation effect, (4) Cloud-agnostic reproduction on three major cloud service platforms includes Azure Cognitive Service \cite{azure-cognitive-services}, Google Cloud Vertex AI \cite{google_cloud_vertex_ai}, and Amazon Web Services Rekognition \cite{aws-ai-services}. 
Our study employs consistency metric ~\cite{paper1} to assess the explanations derived from multiple cloud AI services. 
The experimental results help us observe and discover that data augmentation techniques not only enhance all cloud AI service learning performance but also improve evaluation results from the different XAI methods. 

The adoption of XAI frameworks is designed for data science and machine learning engineers, effectively functioning as a tool for assessment in the development of complex AI systems. A recent study \cite{neghawi2023linking} proposes a multi-level governance pattern that integrates team-level XAI practices with organization-level ethical standards, thereby organizing ethical principles. This work introduces an XAI service framework for AI service practitioners, ensuring alignment with ethical guidelines and organizational values.



The remaining sections are structured as follows: Section \ref{sec:Related Works} explores related works on cloud-based AI services and their explainability challenges. Section \ref{sec:background} summarizes the employed background knowledge. In Section \ref{sec:Post hoc XAI Methods}, we delve into post-hoc XAI methods and their integration into cloud services. Section \ref{sec:Design} presents our microservices-based XAI architecture. Section \ref{sec:provenance} emphasizes the tracing and reproducibility aspects of XAI operations using provenance data.  Section \ref{sec:Discovery} presents the setup and results of the experiment. Section \ref{sec:Further Evaluation} evaluates the XAI service from the system aspect. The paper concludes by summarizing our findings in Section \ref{sec:conclusion}.

\section{Related Works}
\label{sec:Related Works}

This section begins a survey of the growing use of cloud-based AI services for various applications. There is a lack of XAI in cloud services. Following this, a comprehensive overview of various XAI methods and the corresponding implementation frameworks is presented. Lastly, the importance of data provenance within XAI for responsible AI practices is discussed.

\subsection{Cloud-based AI Services}
\label{sec:Cloud-based}
Cloud-based AI services, which offer customized AI models and pre-trained models through APIs for various tasks, have attracted substantial interest due to their versatility and ease of use~\cite{azure-cognitive-services}.
A study~\cite{singh2019comparison} provided a comparative analysis of cloud computer vision services, focusing on their accuracy, performance, and cost. 
However, the study did not delve into the specifics of the AI models used or draw conclusions based on the comparative evaluation.
Image classification uses machine learning algorithms to categorize images based on their content, offering potential applications across various fields.
The lack of explanation can hinder the adoption and trust of these AI systems. 
For instance, a survey~\cite{fabbrizzi2022survey} illuminated the potential biases in visual datasets, emphasizing the necessity for bias discovery and quantification to ensure fairness and transparency in AI solutions.
A recent publication introduces the tool named Threshy \cite{cummaudo2020threshy}, which helps software developers assess an AI service's confidence score. Integrating configuration into client applications and monitoring systems represents an advance in the practical use and safe deployment of AI services.
Both our work and Threshy \cite{cummaudo2020threshy}, aim to assess cloud AI services outputs, but significantly differ in goals and methodologies. While Threshy focuses on decision threshold selection, we concentrate on XAI results from service.

\subsection{The Post-hoc XAI Methods}
\label{sec:reviewXAImethods}
Post-hoc XAI methods are generally classified into two categories~\cite{paper1}: Model-specific and Model-agnostic methods. Model-specific methods, such as CAM-based techniques~\cite{selvaraju2017grad, Grad-cam++, fullgrad, EigenCAM, layercam, fu2020axiom, HiResCAM}, are designed for specific models and require access to key parameters or layer contents of the AI model for generating explanations. In contrast, Model-agnostic methods, such as SHAP~\cite{lundberg2017unified}, are more flexible, capable of producing explanations solely from the input and output of any machine learning model.
The taxonomy uses a tree topology to organize the layers of categories of the XAI methods ~\cite{paper1}. 

\subsection{Frameworks and Packages for Implementing XAI}
\label{sec:framework}
There are several frameworks and packages have been developed to facilitate the implementation of XAI. Dalex~\cite{dalex}, for example, constructs a wrapper around prediction models to enable their exploration and comparison using a multitude of model-level and prediction-level explanations. IBM's Explainability 360 toolkit~\cite{aix360} incorporates an array of model explanation methods within its Python library. Meanwhile, Microsoft's InterpretML~\cite{interpretml} supports eight XAI methodologies. Other libraries such as Captum~\cite{kokhlikyan2020captum} and OmniXAI~\cite{yang2022omnixai} provide extensive collections of XAI techniques. Table \ref{table:xai_frameworks} shows a comparative summary of various XAI frameworks.

\begin{table}[ht]
\centering
\begin{threeparttable}
\caption{Comparative Analysis of XAI Frameworks and Libraries}
\label{table:xai_frameworks}
\footnotesize
\begin{tabularx}{\columnwidth}{>{\hsize=1.2\hsize}X>{\hsize=0.8\hsize}X>{\hsize=1.1\hsize}X>{\hsize=0.8\hsize}X}
\toprule
\textbf{Framework} & \textbf{Supported Data Types} & \textbf{Supported Methods} & \textbf{Cloud Deployment} \\ 
\midrule
Dalex\cite{dalex} & Tab & 1, 2, 3, 4, 5 & N/A \\ 
IBM AIX360\cite{aix360} & Tab/Image/Txt & 1, 2, 6 & Docker \\ 
InterpretML\cite{interpretml} & Tab/Txt & 1, 2, 3, 10 & N/A \\ 
Captum\cite{kokhlikyan2020captum} & Image/Txt & 1, 2, 6, 7, 8 & N/A \\ 
OmniXAI\cite{yang2022omnixai} & Tab/Image/Txt & 1, 2, 3, 4, 5, 6, 7 & N/A \\ 
Vertex XAI\cite{google_cloud_vertex_ai} & Tab/Image/Txt & 1, 2, 8 & GCP service \\ 
XAI Service & Tab/Image/Txt & Encapsulate all methods and frameworks & API-based\\ 
\bottomrule
\end{tabularx}
\begin{tablenotes}
\footnotesize
\item Supported XAI Methods: 1. LIME (Local Interpretable Model-agnostic Explanations)~\cite{ribeiro2016should}, 2. SHAP (SHapley Additive exPlanations)~\cite{lundberg2017unified}, 3. PDP (Partial Dependence Plots)~\cite{friedman2001greedy}, 4. ICE (Individual Conditional Expectation)~\cite{goldstein2015peeking}, 5. ALE (Accumulated Local Effects)~\cite{apley2020visualizing}, 6. LRP (Layer-wise Relevance Propagation)~\cite{binder2016layer}, 7. CAM (Class Activation Mapping)~\cite{selvaraju2017grad,Grad-cam++,EigenCAM,layercam,fu2020axiom,HiResCAM,fullgrad}, 8. Integrated Gradients~\cite{sundararajan2017axiomatic}, 9. Counterfactual Explanations~\cite{wachter2017counterfactual}, 10. Decision Rules~\cite{yang2017scalable}. Note: Tab for Tabular, Txt for Text.
\end{tablenotes}
\end{threeparttable}
\end{table}

Different frameworks and libraries present unique characteristics and orientations. For instance, InterpretML~\cite{interpretml} is primarily designed for tabular and minor text data, while Captum~\cite{kokhlikyan2020captum} is specifically tailored for PyTorch models. Furthermore, these libraries exhibit significant differences in their interfaces, as highlighted by OmniXAI~\cite{yang2022omnixai}. It is also worth noting that comparing the results or performance of these diverse tools across various use cases may prove challenging due to their distinct features and capabilities:
\begin{itemize}
    \item \textbf{Expertise Requirements.} These frameworks require expertise for implementation, potentially limiting their accessibility to a broader user base. A paper \cite{palacio2021xai} discusses the challenges and opportunities of developing user-friendly XAI tools that non-experts can use. 
    \item \textbf{Limited Data and Model Types Support.} The support for various data types and models is limited ~\cite{yang2022omnixai}, restricting their applicability across diverse custom models and databases. 
    \item \textbf{Variation in Method Support.} There is a difference in the methods supported by frameworks. Practitioners encounter difficulties selecting the available methods and performing the necessary data preprocessing.
    \item \textbf{Insufficient Explanation Evaluation.} Many works \cite{lopes2022xai,Zhang2021eva} states that XAI lack standardized evaluation procedures.  
    The current frameworks often lack robust mechanisms for evaluating the explanation results, limiting their effectiveness and potential improvements~\cite{Toward}.
    \item \textbf{Limited Support for Cloud AI Services.} These XAI frameworks rarely support cloud AI services, undermining their utility in multiple cloud platforms. 
    Integrating configuration into client applications and monitoring systems represents an advance in the practical use and safe deployment of AI services.
\end{itemize}

These limitations suggest that the frameworks' inflexibility and extensive preparation may hinder the efficient application of XAI methods.


\subsection{Data Provenance for XAI}
\label{sec:relatedprovenance}

Data provenance in the context of XAI is a component used to trace and reproduce operations. 
A review~\cite{werder2022establishing} explores the ethical considerations and presents the implementation of data provenance to ensure the AI system is responsible. 
Data provenance ensures transparency and accountability~\cite{werder2022establishing} of operations.
The PROV-DM model~\cite{belhajjame2013prov} offers standardized components of representing provenance information. It defines concepts and relationships to capture entities involved in a process, activities that took place, and their interconnections.

Regarding practical implementations, a machine learning pipeline is proposed \cite{samuel2020machine} emphasizing reproducible as a form of data provenance. 
Renku \cite{krieger2021repeatable} is an open online platform tracking data, code, and results with Git. It assists researchers in evaluating, reproducing, and reusing data and algorithms. WholeTale \cite{brinckman2019computing} also promotes reproducibility by enabling researchers to capture and share data, code, and workflow.
The work introduces a system \cite{schelter2017automatically} designed to extract, store, and manage both metadata and provenance information for common artifacts in machine learning experiments, including datasets, models, predictions, evaluations, and training runs. 
The experiment \cite{branco2006enabling} enables provenance to be available as metadata.
This study aims to enable the XAI service system to provide native provenance data.
The XAI operations are reproducible, as provided graph-formatted provenance data, which includes datasets, models, XAI methods, and operational settings.


\section{Background}
\label{sec:background}
In this section, we provide a comprehensive background on the selected XAI techniques and evaluation metrics.

\subsection{CAM-based XAI methods}
\label{sec:CAM-based}

Methods for feature contribution explanation reveal and visualize the correlations between specific content elements, such as pixels in an image, and the resulting decision from a model. 
In explaining vision tasks, Class Activation Mapping (CAM)~\cite{zhou2016learning} initially employs the global average pooling layer to localize features in Convolutional Neural Networks (CNNs). However, the method needs to modify the original layer of the model. As optimizations, Grad-CAM obtains~\cite{selvaraju2017grad} the localization weight by the average gradient of one layer instead of replacing it. Grad-CAM++~\cite{Grad-cam++} is an improved version that uses second-order gradients. EigenCAM~\cite{EigenCAM} takes the first principle component of the activation without class discrimination. LayerCAM~\cite{layercam} spatially weight the activations by positive gradients. XGrad-CAM~\cite{fu2020axiom} scales the gradients by the normalized activations. 

\begin{figure}[ht]
\centerline{\includegraphics[width=0.9\linewidth]{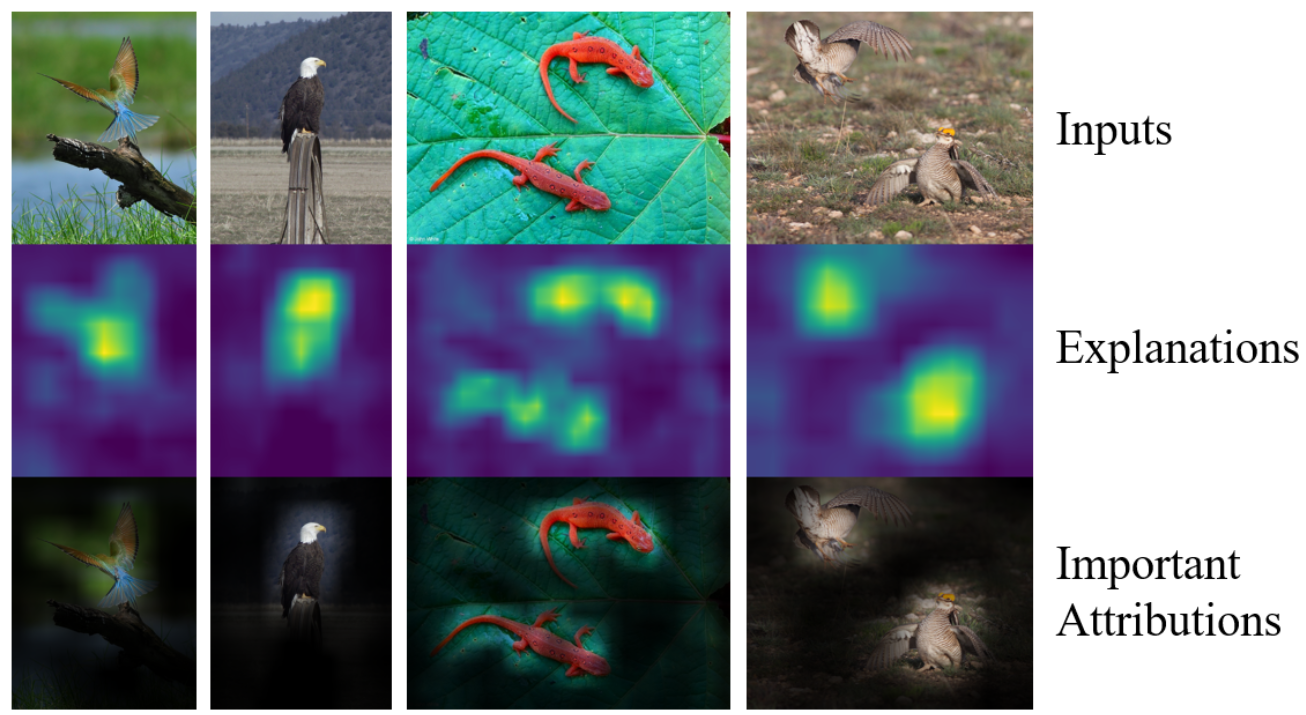}}
\caption{Grad-CAM Visual Explanations on ImageNet Dataset Samples}
\label{Example images}
\end{figure}

The example images of the explanation of the CAM-based method for vision AI tasks are shown in Figure \ref{Example images}. The saliency maps in the second row highlight the regions in the image that contribute to the model prediction. 

\subsection{Metrics for Evaluating XAI Techniques}
\label{sec:Consistency}
The previous work \cite{paper1} reviews and compares XAI evaluation criteria and metrics.
To compare the XAI outcomes objectively, we use consistency metrics \cite{paper1} to perform the assessment and comparison of various XAI methods across identical datasets. We outline the specific equation to compute the intra-consistency \cite{paper1}, also known as the XAI stability.

\subsubsection{The Prediction Change Value}

The metrics are based on the feature importance, also present as feature contribution, provided by XAI methods. 
Assume $\hat{f}(x^{[i]})$ denote the model's predicted value for a dataset $x^{[i]}$. We use $x^{[i]+}$ to symbolize the transformed or augmented data sample, an image under masking. 
We introduce an equation to compute the difference value induced by the feature, denoted as the prediction change value in Equations~\ref{eq:relative}.
\begin{equation}
\label{eq:relative}
\delta{(x^{[i]})}|_{rel} = \left|\frac{ \hat{f}(x^{[i]+}) - \hat{f}(x^{[i]}) }{\hat{f}(x^{[i]})}\right|
\end{equation}
Additionally, the prediction change values of the entire dataset \( X \) can be aggregated. 
The experiment, in Section \ref{sec:scenario2}, uses histograms to present the distribution of prediction changes.

\subsubsection{XAI Consistency Evaluation}
\label{sec:consistency}

The XAI consistency metrics as described \cite{paper1} include both \textit{Intra-Method Consistency} and \textit{Inter-Method Consistency}. In our study, the objective is not to compare inter-XAI methods. Therefore, we specifically focus on \textit{Intra-Method Consistency}, also known as XAI stability evaluation, to assess the quality of the generated explanations by a series of XAI methods. 

Here, we present the stability evaluation. We define the set of explanations, denoted by \( E= \{\xi^{1}, \xi^{2}, \dots, \xi^{m}\} \), where \( \xi^{i} \) represents the \(i\)-th data sample explanation. Next, we define the distance between any two pairs of summaries as \( f_d(\xi^{i},\xi^{j}) \), where \(f_d\) is the distance function, and \( \xi^{i} \) and \( \xi^{j} \) are two different explanation. Then, we describe the combination of choices for selecting any pair of explanations from \(E\), given by \( K=\binom{m}{2} \). This combination represents the number of ways to choose two explanations from the \(m\) in total.

\begin{equation}
f_d^{K} = \frac{1}{K} \sum_{k=1}^{K} f_d(\xi^{i},\xi^{j}), \quad (i \neq j, i \leq m, j \leq m)
\label{eq:stability_metric}
\end{equation}

Here, \(f_d^{K}\) represents the stability metric computed as the average over all \(K\) distance values, where each distance value \(f_d\) is calculated for a specific pair of explanation \( \xi^{i} \) and \( \xi^{j} \).
This metric is used for the assessment of various XAI methods in scenario three, Section \ref{sec:scenario3}. In this case, the distance value \(f_d\) is represented by the prediction change value.

\begin{figure*}[ht]
\centerline{\includegraphics[width=\textwidth]{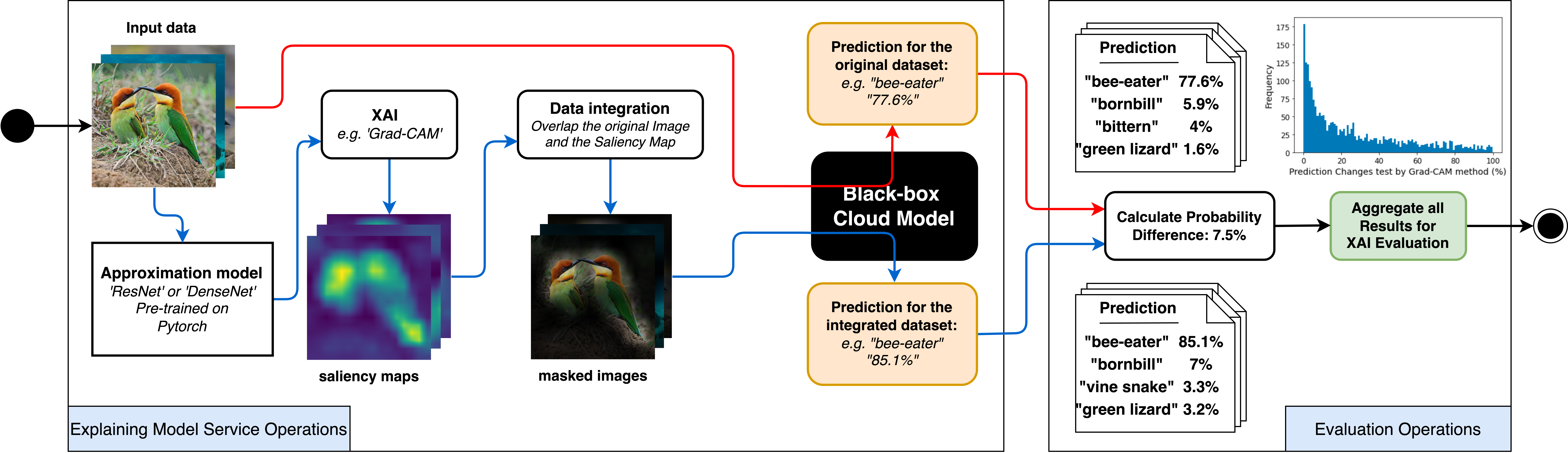}}
\caption{Workflow for Integrating Cloud AI Services with State-of-the-Art Deep Learning Models to Approximate Feature Contribution Values Using Post-Hoc XAI Methods}
\label{flow-chart}
\end{figure*}

\section{The Post hoc XAI Methods with Feature Contribution Explanation}
\label{sec:Post hoc XAI Methods}

In this section, we discuss the post-hoc XAI methods. We explore feature contribution explanation techniques. Then, we present a design that integrates feature contribution XAI methods with cloud AI services, addressing \textbf{\textit{RQ1}}.

\subsection{XAI Methods Taxonomy}
\label{sec:taxonomy}
XAI methods are mainly classified into two categories. One is to develop interpretable models rather than to explain deep learning models~\cite{rudin2019stop}. Interpretable models, such as linear regression, logistic regression, decision trees, and k-nearest algorithms, make predictions in ways that are relatively understandable to humans.
These models are transparent, allowing researchers to understand decision-making by analyzing internal conditions, such as branches in decision trees.
Compared to black-box AI models, interpretable models have the natural advantage of transparency.

The other category pertains to the post-hoc explanation methods~\cite{slack2021reliable}. This approach is typically applied to AI systems that are already deployed and operational, with the need for explanations emerging after the system has made a decision.
Post hoc explanation methods are further grouped into model-agnostic and model-specific. Model-specific methods rely on the structure and properties of the model as information, such as CAM-based techniques~\cite{selvaraju2017grad, Grad-cam++, fullgrad, EigenCAM, layercam, fu2020axiom, HiResCAM}. 
Therefore, model-specific methods typically require a distinct set of parameters and configurations, necessitating the extraction of this data from within the model.
This limitation restricts their applicability across various black-box model scenarios.
On the other hand, model-agnostic methods provide explanations by analyzing the input-output relationship of the AI system without information about the internal structure of the model. For example, SHAP ~\cite{lundberg2017unified} and LIME~\cite{ribeiro2016should}, model-agnostic methods are flexible and can be applied to various black-box models.

\subsection{Integrating Cloud AI with Post hoc XAI}
\label{integratingcloud}

Cloud AI services encapsulate AI models with well-defined RESTful APIs \cite{azure-cognitive-services,google_cloud_vertex_ai,aws-ai-services}. The opacity of the AI models has limitations on the comprehension of how the prediction is made, even though the prediction produces a high accuracy level. It is a limitation for a certain adoption of the AI services that has an emphasis on transparency of the AI models. 

We use post hoc XAI methods for feature influence and causality to devise an approximating model that correlates the inputs and outputs to generate an explanation of feature contribution values. Following this principle, we consider computer vision AI service suitable for adopting post hoc XAI methods for two reasons: (1) A variety of state-of-the-art deep learning models already exist in the field of computer vision as standard models. Even if there is no access to the AI models running within a cloud-based AI service, these established models can serve as suitable approximations; (2) XAI methods specifically designed for elucidating computer vision model explanations are widely available in the literature.

We describe the process and data flow in Figure \ref{flow-chart} at the run-time. 
The initial step involves the preparation of the input dataset. 
Given that cloud model services typically function as a black box, with internal parameters and model structure concealed from the user, we employ approximation models.
The original dataset is then used in three distinct steps: 1) Input images are directly submitted to the cloud-based AI service, which returns their prediction outcomes. 2) The images are processed by an approximation model, employing model-specific XAI methods\cite{selvaraju2017grad,Grad-cam++,EigenCAM,layercam,fu2020axiom,HiResCAM,fullgrad} to generate salience maps. 3) The original images, overlaid with the salience maps, are used to create masked images highlighting only the important features.
Subsequently, both the original and masked images are submitted to the cloud AI service for inference, yielding model predictions and confidence values.
Comparing these two sets of results allows us to quantify the impact of features.
The minimal variance between the two sets indicates similar feature importance values.
We also evaluate the XAI consistency metrics, offering insights into the XAI evaluations within cloud AI services.


%

In summary, the exploration showed that post hoc XAI methods, which elucidate the correlation between inputs and outputs, can approximate feature contributions for cloud AI services.
As illustrated in Figure \ref{flow-chart}, the workflow produces explanations for cloud model predictions.

\section{XAI Service Architecture and API Design}
\label{sec:Design}

\begin{figure*}[ht]
\centerline{\includegraphics[width=\linewidth]{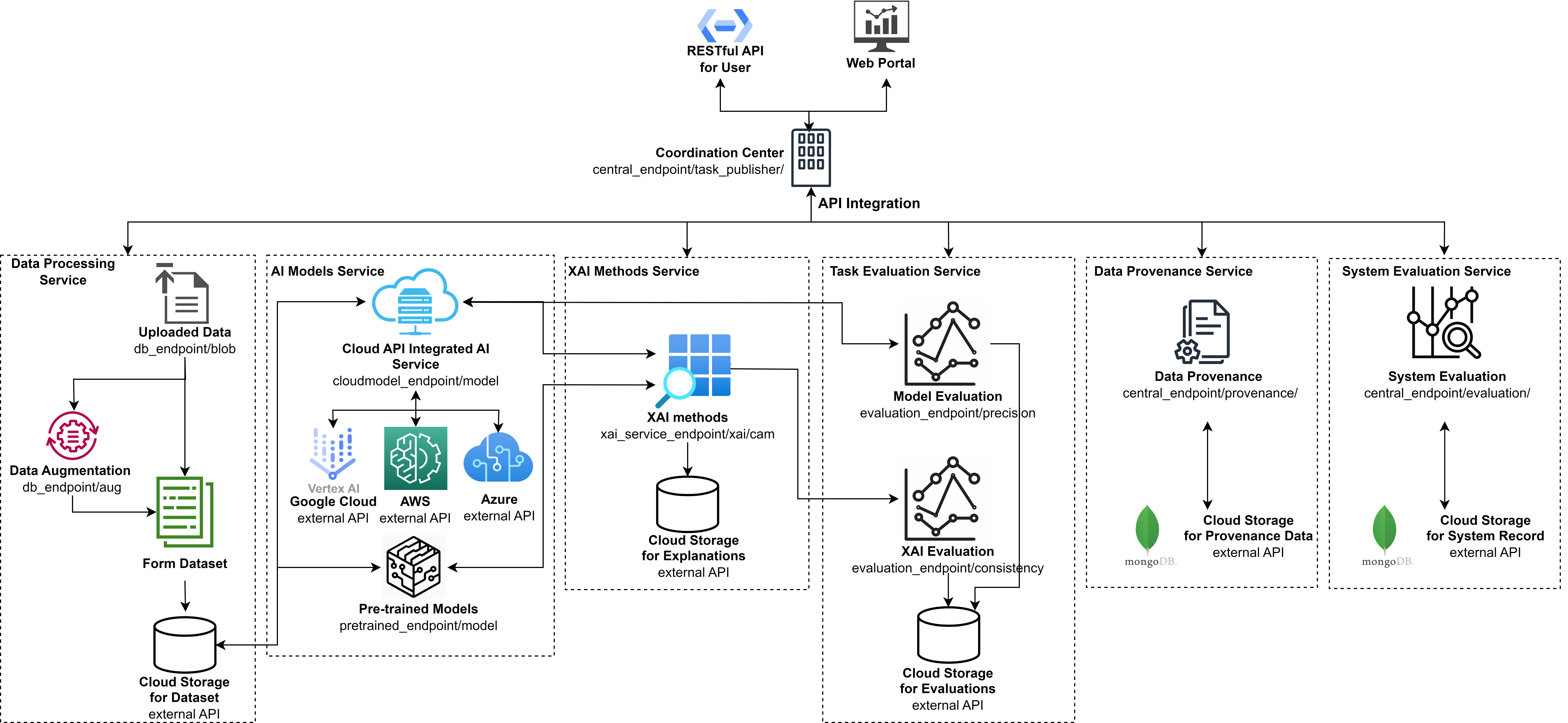}}
\caption{Microservice-Based Reference Architecture for XAI Service}
\label{architecture}
\end{figure*}

We design the XAI operations as well as comprehensive API architecture to answer \textbf{\textit{RQ2}}. This approach led us to define a reference architecture comprising four distinct layers from top to bottom. They are the user interface, the coordination center, the core microservice, and the data persistence. The overview of this reference architecture is illustrated in Figure \ref{architecture}.

\subsection{API Design for Microservices}

One of the key features of this proposed architecture is its non-intrusive application of XAI methods. This approach allows us to apply XAI techniques without significantly disrupting or modifying existing system operations. 
We have completed the API design for this microservice system, with these APIs systematically arranged and hosted on Swagger Hub, adhering to OpenAPI 3.0 standards. Detailed API documentation featuring an array of functional interfaces is available in our published Swagger document.\footnote{\url{https://app.swaggerhub.com/apis-docs/WANGZERUI418/xaiservice/1.0/}}. 

\subsection{API-based Coordination Center}

The coordination center processes requests from the user interface layer and activates the necessary components for task execution.
The core functions of the coordination center are managing microservices, overseeing XAI tasks, and retrieving data.

To manage microservices, the coordination center works with the user interface to handle service registration and modification. It logs endpoints and functional types and initiates the corresponding microservice upon request.

In XAI task management, users submit a task sheet outlining the task type (either XAI or evaluation), associated microservices, dataset, and specific parameters.
The center oversees the entire task lifecycle, using the submitted task sheet as a blueprint.
During execution, the center accesses data, activates the necessary service APIs, and ensures task completion.
Upon task completion, the executor updates the center with the task status, runtime metrics, and provenance data.
Additionally, the coordination center offers a pipeline configuration that sequences task sheets for standard execution of XAI and evaluation tasks.


\subsection{Data Processing Microservice}

Data Processing serves as a critical component in the system architecture. It is primarily responsible for managing data inflow into the system and orchestrating data processing tasks. This microservice takes into account the need for streamlined data management, providing interfaces for users to upload their datasets in specified formats, organize data, and maintain the consistency and integrity of the data.

Data Processing allows users to not only upload their data but also assemble them into datasets. Furthermore, it provides the option to apply data augmentation techniques to enhance the variability and volume of the original data. Data augmentation is instrumental in enhancing the robustness of the AI models and improving the performance of XAI methods. 
The Data Processing Microservice stores data in the user-specified cloud database to ensure data security and ownership. 

This design addresses multiple aspects of data management, including data import, processing, augmentation, and secure storage. The incorporation of these features ensures that the Data Processing Microservice not only maintains the data lifecycle within the system but also provides users with flexibility and control over their own data.

\subsection{AI Model Microservice}

The AI Model Microservice is specifically designed to handle prediction tasks. This microservice is flexible in integrating various cloud AI services from cloud platforms including Amazon Web Services, Google Cloud Platform, and Microsoft Azure. It communicates with these platforms using their specific APIs, ensuring seamless connectivity and usage of the cloud AI services. By connecting with these platforms, users are given access to a broad spectrum of AI models and tools readily available on the cloud.

Another notable feature of the AI Model Microservice is the ability to integrate pre-trained models provided by the users. The integration of user-provided models is facilitated through a well-defined RESTful API specification. Users need to ensure their models adhere to this specification, and once that is done, their models can be easily plugged into the system.

\subsection{XAI Method Microservice}

The XAI Method Microservice in the system's architecture provides explanations for AI model predictions. These XAI methods work based on the contributions of individual features to the final prediction. Once an XAI method is integrated into the system, it can be repeatedly used for various similar types of XAI tasks without the need to consider the development steps again. This approach not only simplifies the procedure of applying XAI methods but also enhances the efficiency of the XAI operations.

Furthermore, this microservice provides a seamless connection with the Data Processing Microservice and the AI Model Microservice through RESTful APIs. This enables it to access data and models, generate explanations, and provide users with comprehensible insights into the AI models.

\subsection{Evaluation Microservice}

The Evaluation in the system brings forward a reliable approach to assess both the AI models and the deployed XAI methods. It is crucial to guide AI practitioners in their decision-making process and enhance trust in the AI models and their explanations.
This service is equipped to evaluate the performance of AI models and XAI methods by examining the explanation results. It employs the consistency evaluation method referenced in \cite{paper1}. This method provides a comprehensive evaluation of the stability and reliability of the XAI methods, thus ensuring that the selected XAI methods yield consistent explanations.

The results derived from this evaluation are efficiently stored and maintained for future reference. This also allows for continuous monitoring and comparison of different models and methods, paving the way for continuous improvement and development in the system. 


In summary, the proposed architecture provides a comprehensive solution for implementing XAI as a service. It consists of several key components: the User Interface, Coordination Center, Data Processing Microservice, AI Model Microservice, XAI Method Microservice, and Evaluation Microservices. These components work synergistically to enable the effective execution and evaluation of XAI tasks. 

\section{Design Provenance Meta Data of XAI Operations}
\label{sec:provenance}

As introduced in \textbf{\textit{RQ3}}, the purpose of provenance data is to trace and reproduce XAI operations. The design of microservices brings another benefit in addition to the integration of heterogeneous components involved in XAI operations. The design helps us understand the relations of these components, focusing on their characteristics with abstraction. For example, we can retrieve the provenance data of two XAI pipelines and identify the difference in the configuration. This helps trace and observe how a certain explanation is produced, improving transparency and trustworthiness.

To this purpose, the provenance data needs to capture the properties of XAI services,  the configuration of XAI tasks and pipelines, an explanation summary from the XAI method and the data set, and XAI cross-validation results.  We apply graph format data to organize, visualize, and understand the relations between XAI provenance data.

\begin{figure}[ht]
\centerline{\includegraphics[width=0.45\textwidth]{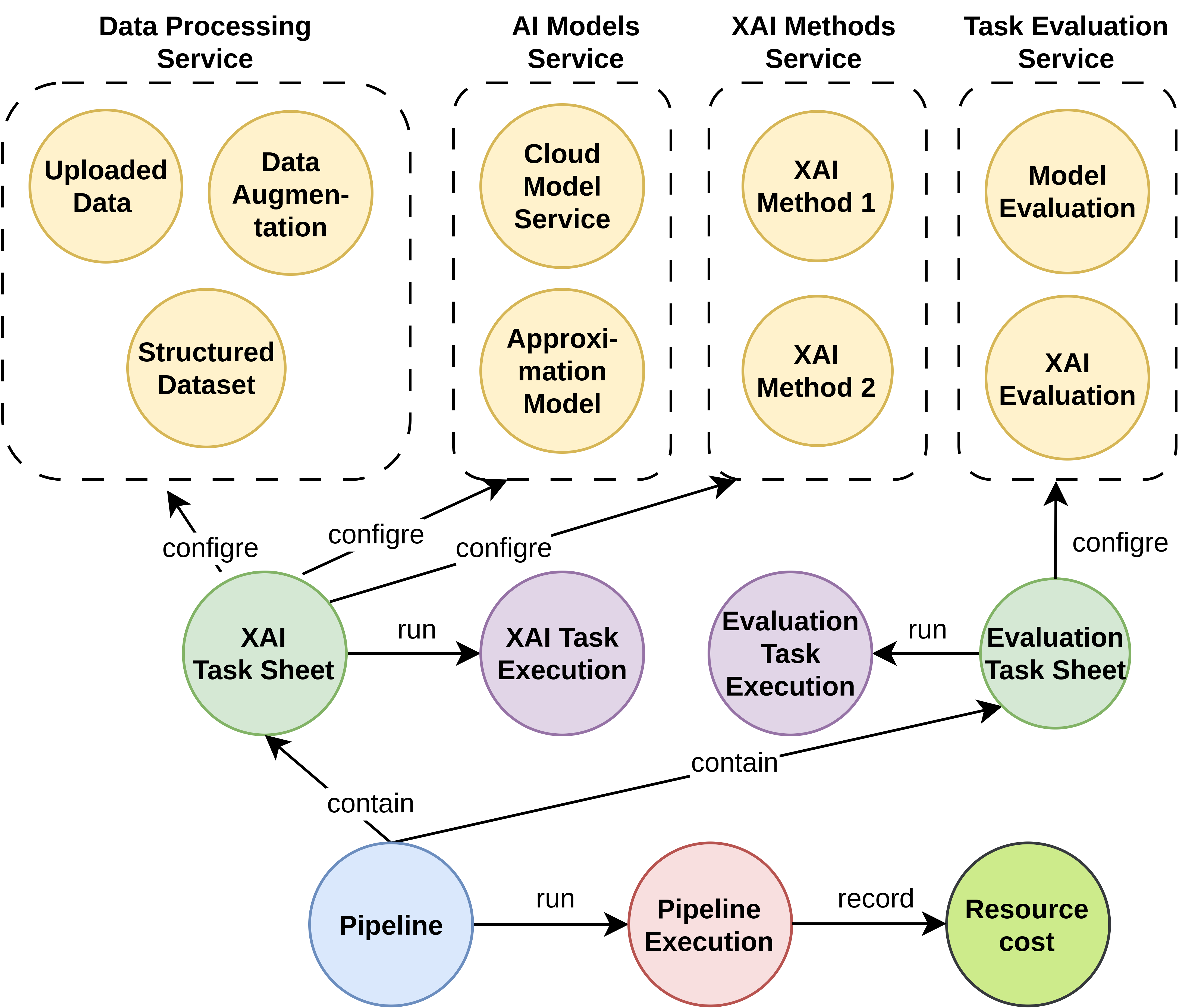}}
\caption{Directed Graph Model of XAI Provenance Data Capturing Microservices, Tasks, and Pipeline Relationships}
\label{Knowledge Graph}
\end{figure}

We model the provenance metadata shown in Figure~\ref{Knowledge Graph}. 
The nodes represent types of XAI provenance data, and the edges represent the relationships between them. 
There are four types of microservices as the yellow nodes in the graph-format provenance data: (1) database, (2) AI, (3) XAI, and (4) evaluation.
These nodes contain the entities that describe the metadata of the microservices, for example, the unique ID, the type, the name, and other note information of the microservice. 

The task sheets are of two types: (1) XAI task sheets and (2) evaluation task sheets. The two task sheet nodes are independent because they deliver different final results, even though they share certain similar components. They record the metadata attributes, including the unique ID, the task type, the task name, the setting parameter, and other detailed configuration information. 
When users execute a task sheet, it generates a unique metadata profile represented as the violet color nodes. These execution nodes capture details including the execution ticket, associated task sheet ID, time of execution, status, results data, and resource consumption.

The pipeline node at the bottom of the graph data typically includes both XAI and evaluation tasks, facilitating comprehensive XAI operations for specific scenarios and settings.
The metadata in the pipeline node comprises the pipeline ID, name, and associated task sheet ID.
Initiating the pipeline generates a pipeline execution node, represented as a distinct node in the graph-structured provenance data.
The pipeline execution node includes log details.
Computational costs are recorded following the execution.

\section{Architecture Usage: XAI-based Cloud AI Service Discovery Scenarios}
\label{sec:Discovery}

The usage of the open API architecture is to obtain observations from discovery scenarios on (1) \texttt{Cross-Cloud AI Service Metrics} - assessing AI services performance variation across multiple cloud services; (2) \texttt{XAI Explanation and Evaluation} - feature contribution explanation and consistency evaluation cross clouds; (3) \texttt{Explainability Improvement} discovering the impact of data augmentation on model explainability across clouds; and (4) \texttt{Operation Provenance} - reproduction of XAI operations. Our scenarios demonstrate the architecture is cloud compatible and thus extensible to develop further discovery scenarios on multiple clouds.

\subsection{The Microservices Configuration and Initialization}

This section presents configuration settings of each microservice in the framework. 
The database microservice uses the ImageNet dataset \cite{deng2009imagenet}, ensuring efficient data storage and access during testing phases.
The model microservice integrates cloud AI services via Restful APIs.
XAI microservices are responsible for generating explanations in saliency maps.
The coordination center effectively manages these diverse microservices, ensuring a streamlined execution of tasks leading to the creation of saliency maps.
Our evaluation task involves validating explanations produced by approximation models. Additionally, we apply data augmentation techniques to explore potential improvements in model and XAI performance.
The source code for implementation of the XAI service discovery scenarios is available in the GitHub Repository \footnote{\url{https://github.com/ZeruiW/XAI-Service}}.

\textbf{Database Microservice.} We implement the database microservice using Azure Blob, a cloud database service that manages data uploading and retrieval for our case studies.

\textbf{Integrate Cloud AI Service.}
Incorporating the APIs of cloud AI models into a microservices infrastructure is crucial for completing XAI service tasks.
Post-hoc XAI methods need access to AI model results in response to various inputs.
This integration provides a unified interface and pipeline for executing XAI experiments.

\textbf{Integrate Pre-trained Models.} We employ ResNet~\cite{resnet} and DenseNet~\cite{huang2017densely} as approximation models due to their exceptional performance on ImageNet image classification benchmarks~\cite{deng2009imagenet}. The pre-trained models are encapsulated within microservices, and their output is standardized to ensure compatibility with other services.

\textbf{XAI Methods Microservice.} 
We interpret image classification models using CAM-based methods such as Grad-CAM~\cite{selvaraju2017grad}, Grad-CAM++\cite{Grad-cam++}, EigenCAM\cite{EigenCAM}, LayerCAM~\cite{layercam}, and XGrad-CAM~\cite{fu2020axiom}.
CAM-based methods generate saliency maps, offering insights into the attention of the convolutional neural network during prediction. These maps highlight key pixels or areas in an image that significantly influence the prediction results of the model.
These methods are encapsulated into microservices, allowing them to download model parameters and dispatch explanations through the coordination center.
We establish an API that facilitates the retrieval of model parameters. This step is critical as it bridges the gap between the XAI and AI services, which are standalone applications. 


\textbf{Evaluation Microservice.} 
We have implemented an evaluation microservice to assess explanations generated by XAI methods, using the previously defined evaluation metrics~\cite{paper1}.
The \texttt{stability}, as detailed in Section \ref{sec:consistency}, forms the backbone of our evaluative procedures.
The evaluation microservice executes the algorithm and forwards the results to the coordination center, offering a holistic view of XAI service performance.
This structure enables a comprehensive assessment of explanations, thereby enhancing the quality assurance of XAI methods in our framework.

\subsection{Scenario One: XAI Consistency Evaluation}
\label{sec:scenario2}
Attribute-based post-hoc XAI methods produce explanations by applying feature masking and mutation, noting changes in predictions.
Evaluating the variation in explanations across different XAI methods is essential to ensure their trustworthiness.
Trustworthiness and transparency are critical technical requirements when applying XAI methods to AI models~\cite{longo2024explainable}.	Improved flow by rephrasing 

As shown in Figure~\ref{flow-chart}, the evaluation consists of three stages, beginning with the preparation of feature changes.
cloud AI services cannot directly derive saliency maps.
To measure changes from feature masking and mutation, we generate saliency maps using approximation models, applying an XAI method.
Next, we overlay the saliency maps onto the original images to create marked images for input into the cloud AI services.
Finally, processing both masked and original images through cloud AI services generates two sets of predictions.
The differences in predictions are analyzed to explain the importance of the features identified by the saliency maps, using XAI.
Figure \ref{flow-chart} demonstrates the workflow to approximate feature contribution exploration with evaluation.
It should be noted that the above inputs are from the test datasets, not from the training dataset. Then, we use the XAI methods and model predictions to compute prediction changes. Finally, we apply consistency metrics to provide the summarized evaluation.

\subsubsection{Selection of Approximation Model}
\begin{figure}[ht]
\centerline{\includegraphics[width=\linewidth]{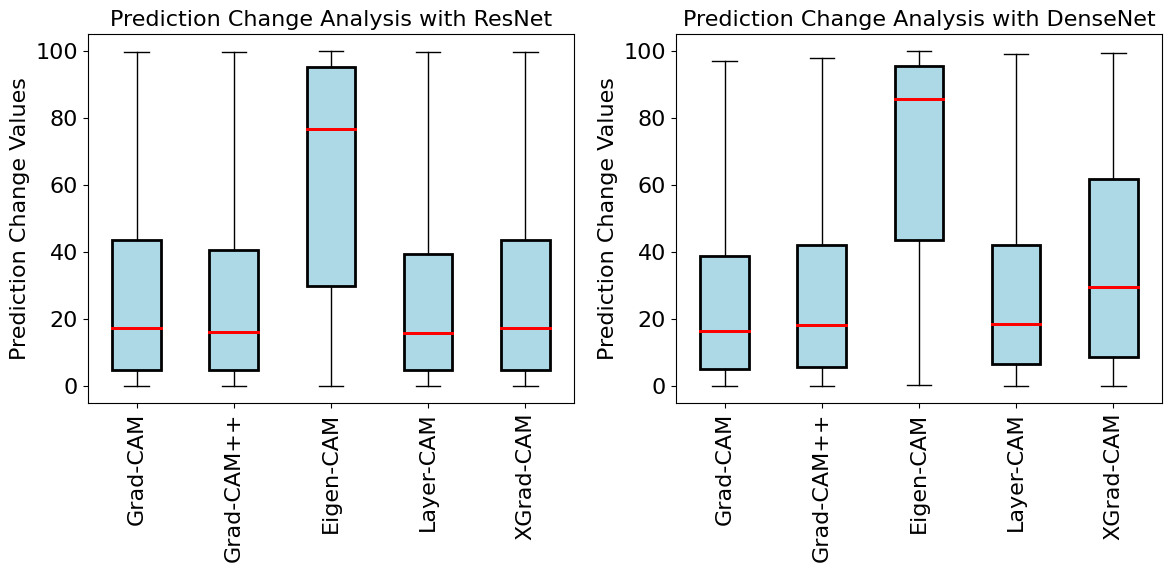}}
\caption{Comparison of Prediction Change Values for ResNet and DenseNet Approximation Models in CAM-based XAI Methods}
\label{ResNetandDensenet}
\end{figure}

In the initial stage, we focus on employing a valid approximation model, as outlined in Figure~\ref{flow-chart}.
We consider two candidate approximation models, ResNet \cite{resnet} and DenseNet \cite{huang2017densely}. 
As a result, Figure~\ref{ResNetandDensenet} illustrates the comparative analysis of prediction change values, highlighting the statistical differences between approximation models. 
Although the differences in prediction change values between ResNet and DenseNet were subtle, ResNet exhibited lower mean change values during the assessment, indicating superior XAI metrics.
Lower computed prediction change values correlate with better XAI consistency metrics.
Therefore, we selected the better option, ResNet, as the approximation model for our subsequent analyses.

\subsubsection{XAI Consistency Evaluation Analysis}
We analyze XAI Consistency Evaluation for Azure Cognitive Services \cite{azure-cognitive-services}, Google Vertex AI \cite{google_cloud_vertex_ai}, and Amazon Rekognition \cite{aws-ai-services}, use CAM-based XAI methods and measure the explainability. 
Subsequently, we aggregate the explanation outcomes across the dataset.



\begin{figure*}[ht]
\centerline{\includegraphics[width=\linewidth]{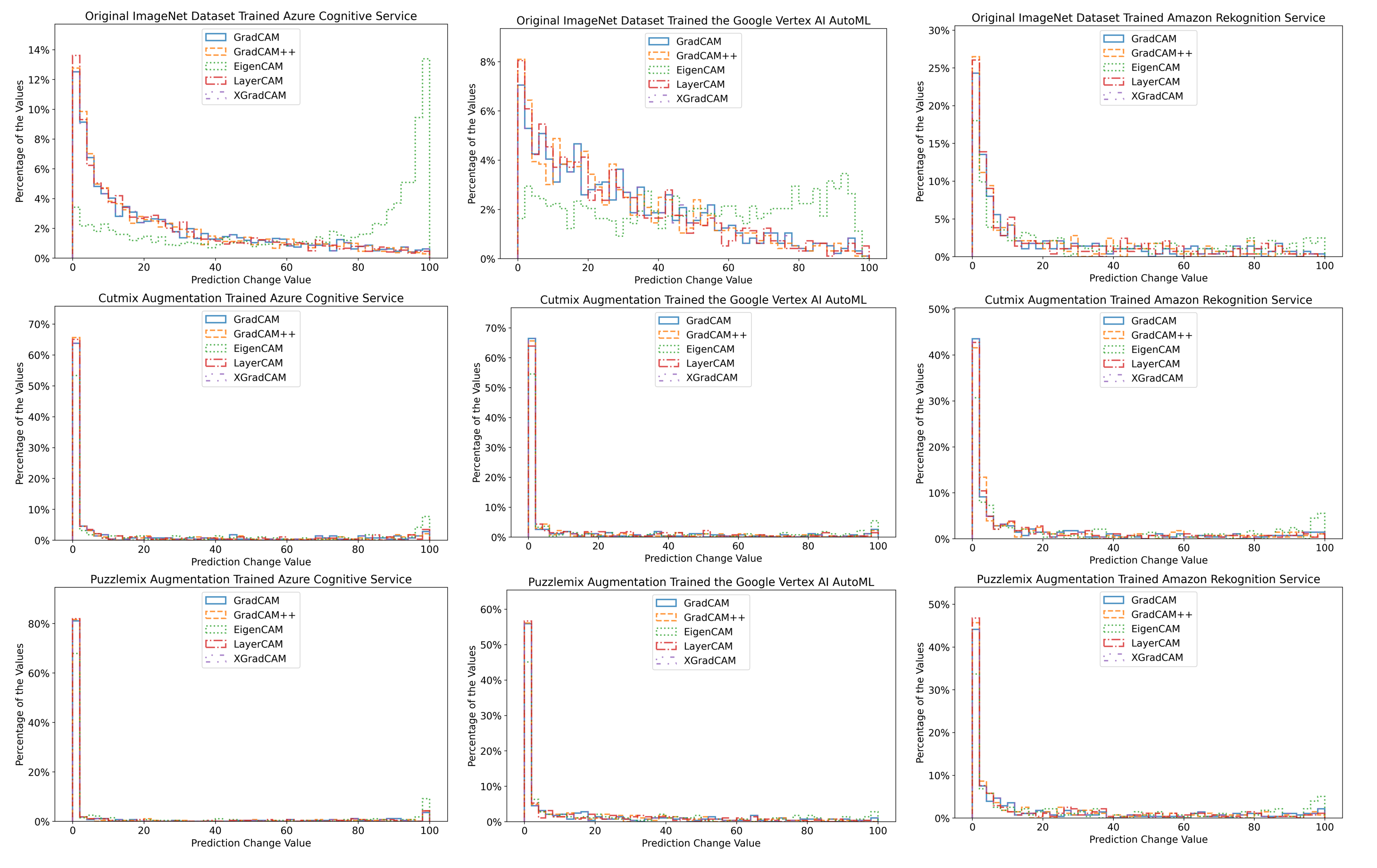}}
\caption{Distribution of Prediction Change Percentages Across Cloud AI Services with Multiple CAM-based XAI Methods}
\label{Histogram}
\end{figure*}

The detailed statistical analysis is presented in the first row of Figure~\ref{Histogram}, where Eigen-CAM displays notably high prediction change values.
The analysis reveals a significant disparity in the performance of EigenCAM compared to other CAM-based methods.
For EigenCAM specifically, 64.7\% of the data showed prediction changes exceeding half, indicating that the explanations provided by EigenCAM do not accurately reflect the model’s predictions.
Among other XAI methods, Layer-CAM performs the best, causing the fewest prediction changes across the test set.
The overall performance of these four XAI methods remains comparable.

\begin{figure*}[h]
\centerline{\includegraphics[width=\linewidth]{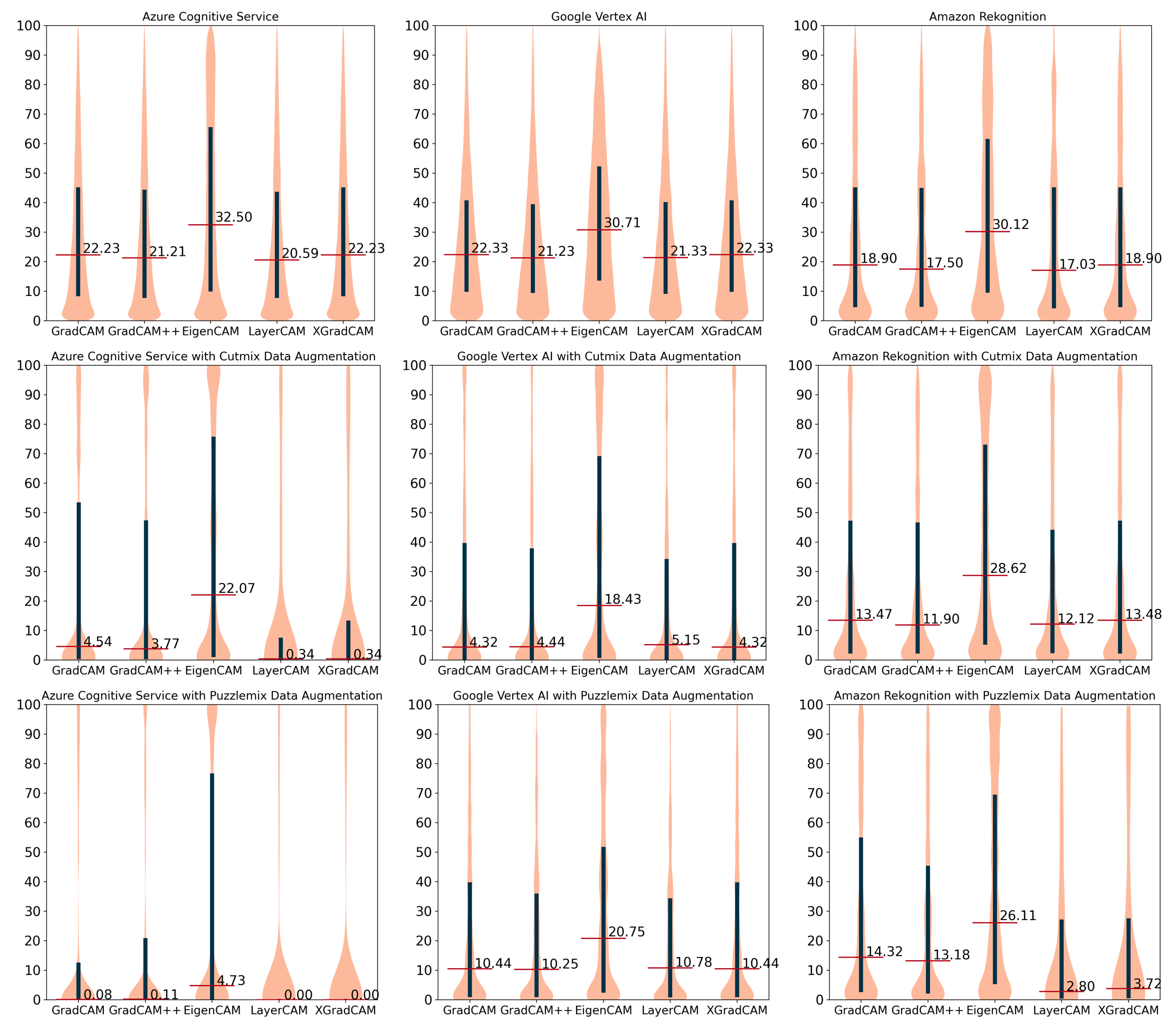}}
\caption{Evaluation of XAI Consistency Across Cloud AI Services Using CAM-based XAI Methods: Lower Mean Values Signify Better Consistency Metric}
\label{stability_cam}
\end{figure*}

For a precise comparison of the prediction change values across various XAI methods, we employed the consistency evaluation Equation \ref{eq:stability_metric}. The results of this assessment are displayed only in the top row of Figure \ref{stability_cam}. This violin plot presents the consistency metrics for explanations derived using five CAM-based XAI methods across leading cloud platforms
A smaller value on the y-axis indicates better consistency evaluation results.
Among these five XAI techniques, both GradCAM++ and LayerCAM show superior consistency metrics. Concurrently, as previously observed, EigenCAM exhibits notably higher values, signifying the worst consistency.

\begin{mdframed}[skipabove=5pt,innertopmargin=5pt,linecolor=black,roundcorner=5pt,backgroundcolor=gray!10] 
Summary - This scenario demonstrates the practical application of the open API architecture, outlined in Section \ref{sec:Design}. 
The findings demonstrate that XAI methods exhibit variability when subjected to the evaluations. 
Notably, GradCAM++ and LayerCAM display superior results, recording the lower values of consistency metrics.
\end{mdframed}

\subsection{Scenario Two:  Data Augmentation Effects on Cloud AI Service Performance}
\label{sec:scenario1}

\begin{table*}[h]
\centering
\caption{Performance Metrics for Cloud-Based AI Services with Different Data Augmentation Strategies}
\begin{tabular}{lccccccccc}
\toprule
& \multicolumn{3}{c}{\textbf{Azure}} & \multicolumn{3}{c}{\textbf{Google}} & \multicolumn{3}{c}{\textbf{AWS}} \\
& Precision & Recall & F1-score & Precision & Recall & F1-score & Precision & Recall & F1-score \\
\midrule
\textbf{No Augmentation}
& \textbf{96.0\%} & 74.6\% & 0.839 & 90.5\% & 41.1\% & 0.565 & 82.8\% & 78.7\% & 0.807 \\
\textbf{Cutmix Augmentation}
& 88.3\% & \textbf{87.0\%} & \textbf{0.864} & \textbf{94.1\%} & \textbf{82.0\%} & \textbf{0.876} & \textbf{84.7\%} & \textbf{82.5\%} & \textbf{0.818} \\
\textbf{Puzzlemix Augmentation}
& \textbf{92.6}\% & \textbf{91.0\%} & \textbf{0.905} & \textbf{93.5\%} & \textbf{81.1\%} & \textbf{0.869} & \textbf{85.3\%} & \textbf{84.0\%} & \textbf{0.828} \\
\bottomrule
\end{tabular}
\label{tab:comparison_augmentation}
\end{table*}

Cloud service platforms such as Microsoft's Azure Cognitive Services \cite{azure-cognitive-services}, Google's Vertex AI \cite{google_cloud_vertex_ai}, and Amazon's Rekognition \cite{aws-ai-services} offer online AI services, including image classification models that can be trained with user-provided datasets. 
Specifically, Azure Cognitive Services limits the number of distinct labels to fifty for image classification. 

Cloud AI services primarily provide evaluation metrics such as Precision, Recall, and F1-Score.
We devise three cases. 
One case uses the original ImageNet dataset without data augmentation, while the other two apply data augmentation to the ImageNet dataset. 
Cloud AI services divide the augmented data into training and test sets.
These services remain as black boxes.

This scenario is to evaluate the effect of the learning performance of these cloud AI services with data augmentation. 
We analyze the Precision, Recall, and F1-score metrics, as detailed in Table \ref{tab:comparison_augmentation}. 
In particular, data augmentation only reduces precision in Azure AI service since the precision is already high without data augmentation. 
However, in all the other cases, both CutMix~\cite{cutmix} and PuzzleMix~\cite{puzzlemix} augmentation algorithms improve the Precision, Recall, and F1-score.
The data augmentation improves the overall cloud model performance F1 score by 0.11 and 0.13.

\begin{mdframed}[skipabove=5pt,innertopmargin=5pt,linecolor=black,roundcorner=5pt,backgroundcolor=gray!10] 
Summary - The use of data augmentation without the expansion of the training dataset reduces the rate of false negatives and improves the F1-score across three cloud AI services on the ImageNet dataset. 
\end{mdframed}

\subsection{Scenario Three:  Data Augmentation Effects on Explanation}
\label{sec:scenario3}

Data augmentation is a tool for boosting the performance of deep learning models by generating a diverse set of synthetic data from existing samples. To further investigate the performance of XAI methods in more complex scenarios, two more advanced data augmentation methods, CutMix \cite{cutmix} and PuzzleMix \cite{puzzlemix}, are introduced and tested in this experiment.
CutMix and PuzzleMix blend images in a patch-wise manner and rearrange patches from multiple images, respectively. They have been demonstrated to enhance model performance significantly~\cite{puzzlemix}. However, the impact of these augmentation methods on the interpretability of models and the effectiveness of XAI methods remains unexplored. 

We commence our exploration by applying the advanced data augmentation techniques, CutMix and PuzzleMix, to our test dataset. The effect of these augmentations on the AI model's performance is summarized in Table \ref{tab:comparison_augmentation}. Upon evaluation of the three cloud AI services, we observe a general improvement in Precision, Recall, and F1-score metrics under the application of CutMix and PuzzleMix, echoing the effectiveness of these methods in enhancing model performance as reported in the literature \cite{puzzlemix}.

The histogram in Figure \ref{Histogram} is the visual representation of prediction changes of the test data set for the three cloud AI services. 
We divide samples into fifty equal bins in the one thousand images test data set to observe their distribution. 
The histogram, Figure \ref{Histogram}, illustrates the XAI prediction change distribution. The middle and bottom rows show that both CutMix and PuzzleMix data augmentation techniques significantly enhance the performance of the model compared to the original dataset. 
The chart shows that the prediction change values for both of these augmentation techniques are smaller than the original dataset, which implies that the XAI methods can generate explanations that are more aligned with the original data. 

This observation supports the claim that these augmentation techniques not only enhance the model's performance but also improve the explainability of the model. It is particularly important in practical applications where both the accuracy of the prediction and model explainability are crucial.

Moreover, the consistency of the XAI methods under these augmentation techniques, as indicated by Figure \ref{stability_cam}, further supports this observation. Consistency here refers to the stability of the XAI methods in providing explanations for more data samples. From the figure, it can be seen that the XAI methods tend to have better consistency with CutMix and PuzzleMix augmentation compared to the original dataset. This result discovered that these augmentation techniques could potentially enhance XAI results. 

Combined with the results in Table \ref{tab:comparison_augmentation}. An analysis of traditional performance metrics such as precision, recall, and F1 score might suggest a minimal effect of the CutMix and PuzzleMix data augmentation techniques on model performance. 
However, this observation could potentially underestimate the broader influence of these augmentation techniques. 
When examining the outputs of the XAI methods, these augmentation techniques demonstrate a pronounced enhancement. 

Furthermore, through the observation of the histogram Figure \ref{Histogram}, we notice a marked improvement in the performance of the EigenCAM method after the application of data augmentation. Initially, EigenCAM seemed to provide less consistent explanations for the original model, as evidenced by the large prediction change values and bad consistency. This suggested a significant discrepancy between the identified salient features and the model's actual decision-making focus. 

However, upon applying the data augmentation techniques, similar to other applied XAI methods, EigenCAM demonstrated a marked increase in consistency and accuracy of the provided explanations. This observation underlines the potential of data augmentation techniques to significantly enhance the performance of XAI methods that initially appear suboptimal. 
This discovery highlights the need to consider data augmentation as a valuable tool not only for enhancing model performance but also for optimizing the explainability provided by XAI techniques.

The usage of XAI, as illustrated in Figures \ref{Histogram} and \ref{stability_cam}, unveils an alternative dimension to this assessment. Through the lens of XAI, a considerable enhancement in model explainability is discernible with the data augmentation techniques. This finding makes it clear that even though data augmentation techniques might not significantly shift conventional performance metrics, they notably enhance a model's explainability. This insight enriches our understanding of the model's decision-making process.

Our discovery highlights the importance of comprehensive evaluation strategies. Relying solely on traditional performance metrics might overlook these techniques' crucial benefits. 
In Figure \ref{stability_cam}. The violin plot's highlight area represents the frequency distribution of defined distance values, and the vertical bars represent the range of the distance value from the first to the third quartile. The median values are marked and indicated. 

Among the specific XAI methods, the Layer-CAM method shows the best consistency since it has the smallest medium and mean values, followed by Grad-CAM++, Grad-CAM, and XGrad-CAM, which are slightly less stable but still comparable with LayerCAM. 
Eigen-CAM shows poor consistency, and its overall prediction changes value distribution is more elevated than the other XAI methods. 

\begin{mdframed}[skipabove=10pt,innertopmargin=5pt,linecolor=black,roundcorner=5pt,backgroundcolor=gray!10]
Summary - The scenario concludes that data augmentation techniques can enhance explanations from cloud AI services without altering the model's structure. XAI consistency values improve by 12.79 and 14.10 respectively. 
\end{mdframed}

\subsection{Scenario Four: Reproduction of XAI Explanation}
\label{sec:scenario4}
\begin{figure}[ht]
\centerline{\includegraphics[width=\linewidth]{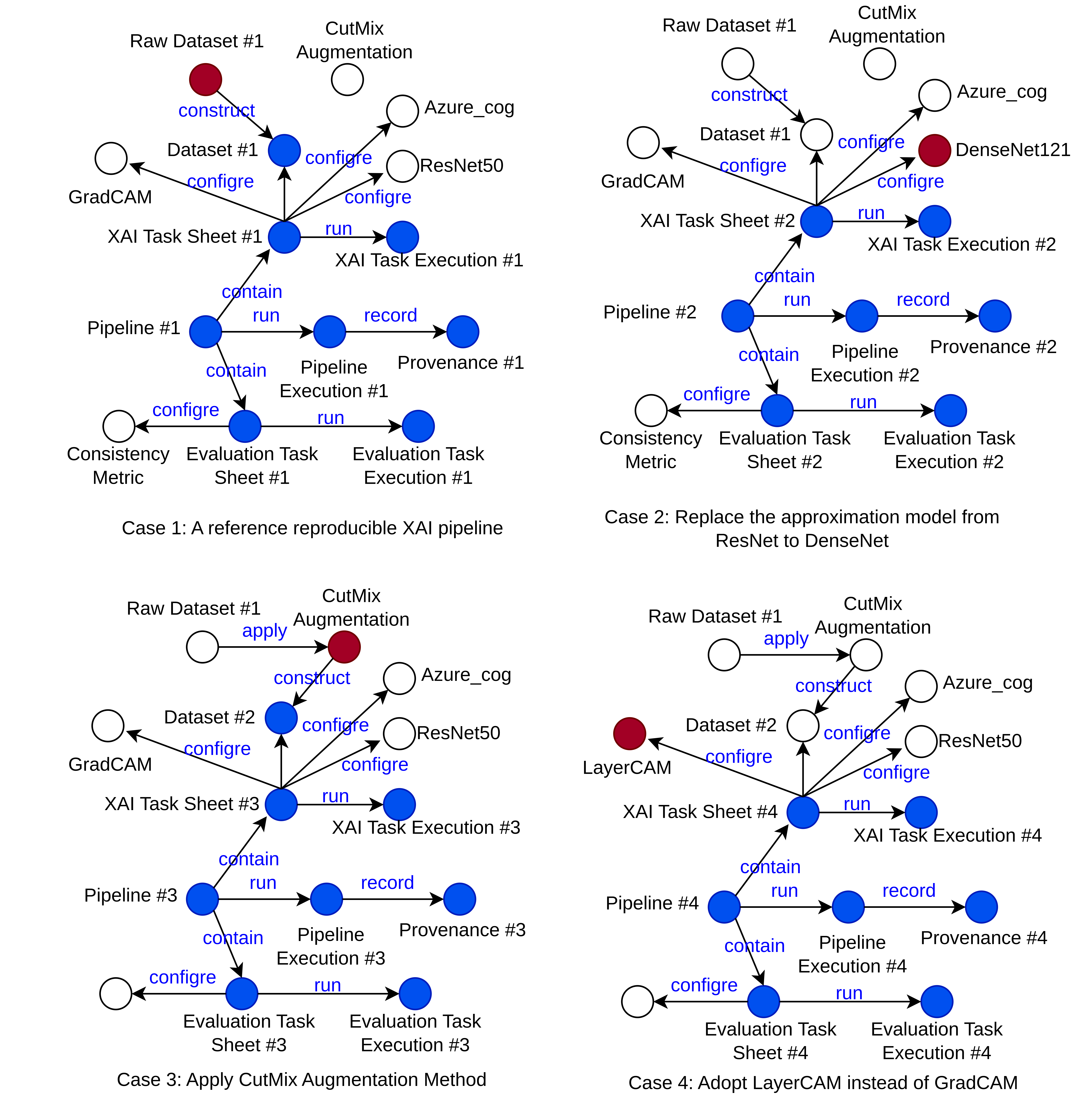}}
\caption{Provenance Graph by XAI Service. Operations using ResNet and Azure Cognitive Service with Grad-CAM, and explanation evaluation}
\label{Provenance Graph example}
\end{figure}

From the scenarios described above, we have experienced the complexity of XAI evaluation.
XAI evaluation extends beyond merely running a single XAI algorithm.
It involves a complex composition of entities and tasks, including datasets, feature masking and mutation, data augmentation, benchmark model approximation, cloud AI services, and various XAI methods.
The scale of these combinations can rapidly escalate the complexity involved in providing explanations.
Tracing XAI pipelines and ensuring transparency in XAI explanations are key technical requirements for XAI evaluations. We devise four cases to demonstrate the XAI pipeline provenance in our framework design aspects. Figure~\ref{Provenance Graph example} illustrates the provenance graph of XAI variation using ResNet, Azure Cognitive Service, and Grad-CAM XAI method.  The metadata defined in section~\ref{sec:provenance} become the vertices in the graph. Each edge in the provenance graph is an API-based service invocation. A red colored vertex indicates a certain variation introduced to the configuration of a pipeline. The blue vertices indicate related vertices in the paths that are affected by the change. A variation results in a new pipeline graph that represents the adjustments.

\textbf{Cross-dataset adaptability.} Database microservice is designed to serve data via a RESTful API, integrating with cloud data storage services, in the case study, Azure Blob Storage. It ensures the data required by XAI are organized by group and ready for retrieval. By altering the configuration parameters, users can invoke alternative cloud databases, enhancing the system's flexibility.  Case one shows a reference pipeline configuration. The dataset can be replaced without changing other configurations in a pipeline. 

\textbf{Configurable model integration} 
Case two illustrates the necessity for users to replace the model currently in use, exemplified by replacing the ResNet model with DenseNet. As indicated by a red node in the graph data, users have the flexibility to configure the model. 
This modification necessitates the re-execution of the entire pipeline, resulting in changes to the content associated with the blue nodes. The graph provenance data captures and documents user interactions and the resultant outcomes, providing a detailed account of the effects of modifications.
 
\textbf{Incorporation of data augmentation techniques.} Datasets are labelled with distinct names and group identifiers. This feature is used to generate augmented datasets by specifying the original dataset and the augmentation method. The resulting datasets can then be employed to retrain AI models, providing enhanced model performance. Case three in Figure~\ref{Provenance Graph example} shows the pipeline's variation with data augmentation applied. Compared to case one and case two, the vertex \textit{CutMix Augmentation} is connected to a new dataset generated \textit{DataSet \#2}.

\textbf{Extensibility to XAI methods.} For attribute-based XAI methods, the explanation varies by XAI methods for the same dataset and the same model~\cite{wang2024xaiport}. With well-structured API interfaces, new XAI methods can be integrated by encapsulating the input and output conformable to open API specifications.  Case four adopts a different XAI method \textit{LayerCAM} in the pipeline compared to Case Three.

XAI task reproducibility offers significant benefits in the domain of services.
First, it increases the transparency of the service by offering a clear understanding of the steps leading to a specific result. 
The heightened transparency, in turn, builds user trust as they can independently verify the results.
Secondly, the service's adaptability is enhanced. Users can adjust models or configurations based on previously successful operations.
Thirdly, reproducibility facilitates auditing, which can lead to further improvements in service reliability and quality.
Provenance data is persistently stored in the cloud-based, user-configured database.
This ensures that the user-owned provenance data accurately represents the XAI service.

\begin{mdframed}[skipabove=10pt,innertopmargin=5pt,linecolor=black,roundcorner=5pt,backgroundcolor=gray!10]
Summary -  The XAI service is composition of diverse cloud databases, data augmentation, and XAI methods. The provenance of XAI pipelines enables reproducible XAI processes by re-running the configuration of the graph-format provenance data. Any changes are tractable from the pipeline's configuration which is an encapsulation at the microservice level. 
\end{mdframed}

\subsection{The Discover Scenarios Conclusion}
\label{sec:scenarioconclusion}

In conclusion, the scenarios we explored manifest the effectiveness and versatility of our proposed XAI microservice architecture. This structure is instrumental in facilitating the evaluation and comparison of various XAI techniques and cloud-based AI services. 
We summarize the observations:

\begin{itemize}
\item Both CutMix and PuzzleMix data augmentation techniques have been found to significantly improve the performance metrics (Precision, Recall, and F1-score) of cloud-based AI services, as demonstrated in Table \ref{tab:comparison_augmentation}.
\item The augmentation techniques have been shown to reduce prediction change values, representing an improvement of XAI performance, as depicted in Figure \ref{Histogram}. 
\item Traditional performance metrics may understate the broader benefits of data augmentation techniques. By considering XAI methods in evaluation, we reveal an enhanced explainability of AI models, indicating the augmentation techniques' potential to create more trustworthy AI systems. 
\item There is an observable improvement in the consistency of XAI methods under the CutMix and PuzzleMix data augmentation techniques, as indicated by Figure \ref{stability_cam}. This underscores the potential of these techniques in enhancing XAI results consistency and improving model performance.
\item Among the selected XAI methods, the consistency of Layer-CAM outperforms other XAI methods when applied to Azure Cognitive Service using ResNet, as demonstrated in Figure \ref{stability_cam}. Conversely, Eigen-CAM exhibits poor consistency and a higher distribution of prediction change values, indicating lesser reliability.
\end{itemize}

These insights reinforce the importance of incorporating XAI methods in the evaluation process to understand the impact of data augmentation techniques, thereby broadening our perspective on AI model performance.

\section{Quantitative Evaluation of XAI Service Performance}
\label{sec:Further Evaluation}

Upon completion of the experiments, we delve further into the evaluation of our XAI service, examining implementation complexity, computational resource usage, and the reproducibility of results. These elements play a crucial role in determining the effectiveness of any service system.

\subsection{The Effort Required for Microservice Development}

To evaluate the effort involved in developing the added microservices, we quantified the Lines of Code (LoC) needed for integration. Specifically, we focus on the LoC necessary to incorporate a given model or algorithm into a microservice architecture using a provided template. Table~\ref{loc} presents the statistical breakdown of this effort.

\begin{table}[htbp]
\caption{Quantitative Analysis of LoC for Algorithm Integration into Different Microservices}
\renewcommand{\arraystretch}{1.2}
\begin{center}
\begin{tabular}{M{1.0cm} | M{1.4cm} M{1.4cm} M{1.4cm} M{1.4cm}}
\hline

 & \textbf{The Template} &\textbf{Azure Cognitive Service} &  \textbf{Google Vertex AI AutoML} &  \textbf{Amazon Rekognition}  \\ 
\hline
 \textbf{LoC}& 76 &   144  &     198  &   148  \\

\hline
\end{tabular}
\label{loc}
\end{center}
\end{table}

To perform the migration of pre-existing code into the microservice architecture, the development process involves several key steps:

\begin{enumerate}
\item Initialize an HTTP server framework as the microservice's base.
\item Incorporate the pre-existing code into this server framework as a modular function.
\item Standardize the input: Convert HTTP request data into a format suitable for the model or function.
\item Standardize the output: Transform the model or function output into a proper HTTP response and, if necessary, update the database with the result.
\end{enumerate}

By outlining the development process in this way, we provide a roadmap for developers who seek to transition from monolithic or less modular architectures into a microservices framework.

\subsection{Integration of Methods from Other XAI Frameworks}

\begin{figure}[ht]
\centerline{\includegraphics[width=\linewidth]{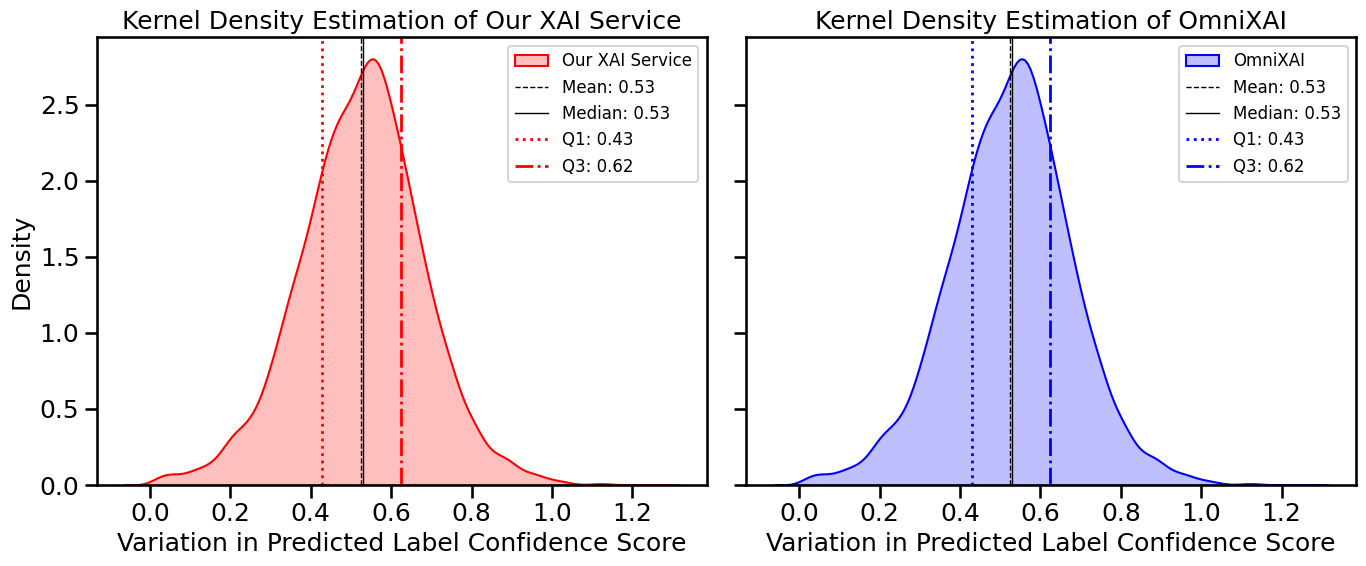}}
\caption{Kernel Density Estimation of prediction confidence changes for 'Our XAI Service' vs. 'OmniXAI'. }
\label{compareomnixai}
\end{figure}

We perform a comparative analysis to showcase the external XAI tool's integration capabilities within our XAI service, with a focus on incorporating the GradCAM \cite{selvaraju2017grad} method.
In Figure \ref{compareomnixai}, the comparison utilizes ten thousand images from the ImageNet dataset \cite{deng2009imagenet} input to the ResNet model \cite{resnet}.
The Kernel Density Estimation (KDE) plots show a consistent prediction confidence distribution between our service and the standalone algorithm from OmniXAI.
Our XAI service is effective in streamlining complex XAI methods via automated pipelines, simplifying the user experience with the same algorithm executed.
Our service, by offering integration and RESTful API, makes advanced XAI techniques accessible for explaining model predictions with less operations.




\subsection{Evaluation of Computational Resources}
\label{sec:Overhead}

The concept of computational overhead extends to the resources required for computation on both the server and client sides. Cloud-based AI services abstract away resource management complexities, charging users based on API request count or usage duration.
To provide a comprehensive analysis, we measured computational resources utilized by different XAI methods. This is particularly pertinent for applications that demand real-time responsiveness and for those processing large datasets.
We maintained a consistent infrastructure across all XAI methods to ensure a fair comparison. Data related to microservice response time and computational resource utilization were collected and analyzed. Through code audits and local tests, we found that resource usage scales linearly with the workload.
Table \ref{EnergyConsumption} presents a comparative view of the average energy utilization for different XAI methods and their corresponding microservices on a per-data-sample basis.

\begin{table}[ht]
\caption{Comparative Energy Consumption of Microservices (Average per Data Sample, in $10^{-6}$ kWh)}
\renewcommand{\arraystretch}{1.2}
\begin{center}
\begin{tabular}{M{2cm} | M{1cm} M{1cm} M{1cm} M{1.5cm}}
\hline
\textbf{Microservices} & \textbf{\textit{CPU Energy }} &  \textbf{\textit{GPU Energy}} &  \textbf{\textit{RAM Energy}} &  \textbf{\textit{Energy Consumed}} \\ 
\hline
\textbf{Data process} &  3.60 & N/A & 0.08 & 3.68 \\
\textbf{Resnet model} &  1.61 & 3.77 & 0.02 & 5.39 \\
\textbf{Densenet model} &  0.38 & 0.89 & 0.02 & 1.28 \\
\textbf{GradCAM } &  8.32 & 19.50 & 0.10 & 27.91 \\
\textbf{GradCAM++ } &  7.09 & 15.93 & 0.09 & 23.11 \\
\textbf{EigenCAM } &  121.48 & 107.54 & 1.71 & 230.74 \\
\textbf{LayerCAM } &  7.21 & 10.08 & 0.09 & 17.37 \\
\textbf{XGradCAM } &  6.77 & 17.35 & 0.08 & 24.20 \\
\textbf{Evaluation} &  16.32 & N/A & 0.17 & 16.49 \\
\hline
\end{tabular}
\end{center}
\label{EnergyConsumption}
\end{table}

From Table \ref{EnergyConsumption}, we observe that Layer-CAM has the lowest computational overhead. This suggests that Layer-CAM might be better suited for systems with limited compute resource availability.
On the other hand, Eigen-CAM shows the highest computational resource utilization and execution time, in line with its lower performance in the prediction change and stability tests.

Overall, these results imply the cost for the quality of the explanations, computational efficiency, and time efficiency. This indicates that the choice of the XAI method should consider the specific requirements and constraints of the application.

\section{Conclusion}
\label{sec:conclusion}

The growth of cloud-based AI brings the urgent need for explanation in these services. 
Our study responds to this necessity by introducing an XAI service architecture.
Utilizing a microservices design, we facilitate the flexible deployment and evaluation of XAI methods for both pre-trained and cloud-based AI models.
We present a workflow that retrieves and evaluates feature contribution explanations for cloud model service.
We rigorously evaluated our service in computer vision scenarios.
Importantly, the incorporation of data augmentation techniques such as CutMix and PuzzleMix not only increases the performance of the underlying models but also enhances the quality of these explanations. 
Our detailed assessment of computational efficiency highlights the architecture's broad applicability across different scenarios and cloud platforms. 
A key feature of our service is its provenance data design, making all XAI operations traceable and reproducible. 
This work provides a concrete framework that provides automated XAI to cloud AI services. 
The principles and design of our architecture offer a reference for future developments in the adoption of XAI operations.




\bibliographystyle{IEEEtran}
\bibliography{Reference}
\vspace{-2cm}
\begin{IEEEbiography}[{\includegraphics[width=1in,height=1.25in,clip]{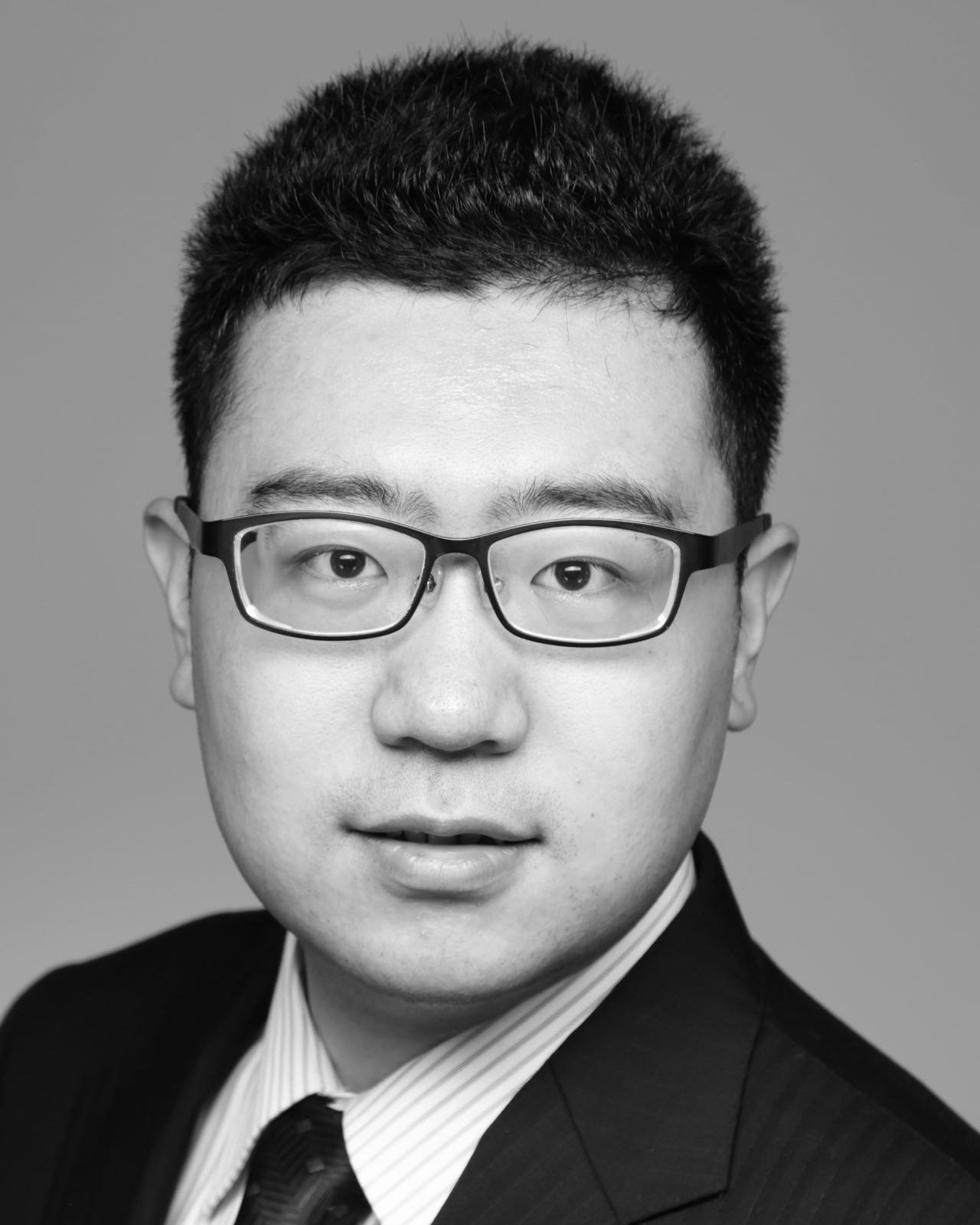}}]{Zerui Wang}is a Ph.D. student in the Department of Electrical and Computer Engineering at Concordia University. Zerui's research interests include machine learning, explainable AI, and cloud computing. Contact him at zerui.wang@mail.concordia.ca.
\end{IEEEbiography}
\vspace{-2cm}
\begin{IEEEbiography}[{\includegraphics[width=1in,height=1.25in,clip,keepaspectratio]{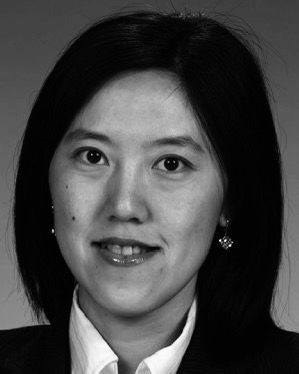}}]{Dr. Yan Liu} is a tenured Associated Professor and Gina Cody Research and Innovation Fellow at Concordia University. Before the faculty position, Yan worked as a Senior Research Scientist at the National ICT Australia (NICTA) laboratory and US Department of Energy Pacific Northwest National Laboratory with ten years of experience with large-scale software systems. As a tenured faculty, Yan's research is generously funded by NSERC Discovery Grants, Quebec FRQNT New Research Award, and MITACS and industry collaborators in the domains of telecommunication, health care, senor networks, NLP for public services, cloud game servers, and digitization of building architecture design. Yan has two US patents granted. Her recent work is defining an evaluation framework for explanation consistency. Contact her at yan.liu@concordia.ca
\end{IEEEbiography}
\vspace{-2cm}
\begin{IEEEbiography}[{\includegraphics[width=1in,height=1.25in,clip,keepaspectratio]{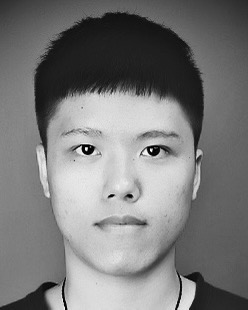}}]{Jun Huang} is a Ph.D. student in the Electrical and Computer Engineering department at Concordia University. His research interests focus on eXplainable AI in Convolution Neural Networks. He is a graduate student member of IEEE. Contact him at jun.huang@mail.concordia.ca
\end{IEEEbiography}

\end{document}